# On the thermodynamical analogy in spin-polarized density functional theory

T. Gál[1,*], P. W. Ayers[2], F. De Proft[1], P. Geerlings[1]

[1]General Chemistry Department (ALGC), Free University of Brussels (VUB), 1050 Brussel, Belgium

[2]Department of Chemistry, McMaster Univeristy, Hamilton, Ontario, Canada L8S 4M1

**Abstract:** The thermodynamical analogy of density functional theory, which is an organic part of the spin-independent version of the theory, is reconsidered for its spin-polarized generalization in view of the recently uncovered nonuniqueness of the external magnetic field $B(\vec{r})$ corresponding to a given pair of density $n(\vec{r})$ and spin density $n_s(\vec{r})$. For ground states, the nonuniqueness of $B(\vec{r})$ implies the nondifferentiability of the energy functional $E_{v,B}[n,n_s]$ with respect to $n_s(\vec{r})$. It is shown, on the other hand, that this nonuniqueness allows the existence of the one-sided derivatives of $E_{v,B}[n,n_s]$ with respect to $n_s(\vec{r})$. Although the $N$-electron ground state can always be obtained from the minimization of $E_{v,B}[n,n_s]$ without any constraint on the spin number $N_s = \int n_s(\vec{r})d\vec{r}$, the Lagrange multiplier $\mu_s$ associated with the fixation of $N_s$ does not vanish even for ground states. Rather, $\mu_s$ is identified as the left- or right-side derivative of the total energy with respect to $N_s$. This justifies the interpretation of $\mu_s$ as a (spin) chemical potential, which is the cornerstone of the thermodynamical analogy.

*Email address: galt@phys.unideb.hu

# I. Introduction

Density functional theory (DFT) [1] is not only one of today's most powerful computational tools in quantum chemistry and solid-state physics, but it provides a natural framework for the definition of a wide range of chemical reactivity descriptors. This subfield of DFT [2-5] is usually called conceptual DFT. DFT's power lies in the use of the electron density $n(\vec{r})$ as the basic variable in the description of electron systems in a scalar external potential $v(\vec{r})$, instead of the complicated many-electron wavefunction $\psi(\vec{r}_1 s_1, ..., \vec{r}_N s_N)$. DFT has been extended to embrace electronic systems in a magnetic external field $\vec{B}(\vec{r})$ as well. The most elementary extension is the so-called spin-polarized density functional theory (SDFT), where magnetic fields act only on the spins of the electrons, and the dipolar interaction between spins is excluded [1,6,7]. In SDFT, in addition to the electron (number) density $n(\vec{r})$, the magnetization density $\vec{m}(\vec{r})$ enters as basic variable. Functionals $A[n]$ are replaced by functionals $A[n, \vec{m}]$; most importantly, an energy functional $E_{v,\vec{B}}[n, \vec{m}]$ appears in the place of the original $E_v[n]$. SDFT is useful even in the absence of magnetic fields; for example, SDFT allows one to extend conventional DFT beyond ground states and treat the lowest-energy state of every spin multiplicity. For such applications, a simplification of SDFT, emerging in the case of collinear magnetic fields, suffices: the first two components of the magnetization density can be taken to be zero, leaving only $m_z(\vec{r})$ as the other (scalar) variable beside $n(\vec{r})$. Instead of $m_z(\vec{r})$, then usually the so-called spin (polarization) density, $n_s(\vec{r}) = -(1/\beta_e) m_z(\vec{r})$, is used, because of its simpler connection to the spin-up and spin-down components of $n(\vec{r})$: $n_s(\vec{r}) = n_\uparrow(\vec{r}) - n_\downarrow(\vec{r})$.

Recently, much attention has been focused on the SDFT generalization of the DFT-based chemical reactivity descriptors [8-29], which have proved to be very useful in the spin-independent version of DFT [2,3]. However, the finding of Eschrig and Pickett [30] and Capelle and Vignale [31] that the correspondence between $\left(n(\vec{r}), m_z(\vec{r})\right)$ [or $\left(n_\uparrow(\vec{r}), n_\downarrow(\vec{r})\right)$] and $\left(v(\vec{r}), B(\vec{r})\right)$ is not one-to-one for nondegenerate ground states [32], but $B(\vec{r})$ is determined by $\left(n(\vec{r}), m_z(\vec{r})\right)$ only up to a nontrivial additive constant [30] (see also [33-38]), is a threat to the SDFT generalization of conceptual DFT. Specifically, problems may arise because the nonuniqueness of $B(\vec{r})$, for ground states, implies the nonexistence of the derivative of the SDFT energy density functional $E_{v,B}[n, n_s]$ with respect to $n_s(\vec{r})$, and many of the reactivity indices are linked to that derivative. In this paper, it will be shown that although the nonuniqueness of $B(\bar{r})$ excludes the existence of the full derivative of $E_{v,B}[n,n_s]$ with respect to $n_s(\bar{r})$ for ground states, it does not exclude the existence of the one-sided derivatives. Therefore the definition of one-sided reactivity indices remains possible.

## II. Existence of $F[n,n_s]$'s derivative over spin density domains of constant spin number

The nonexistence of the derivative of $E_{v,B}[n,n_s]$ with respect to $n_s(\bar{r})$ is due to the following. The energy functional $E_{v,B}[n,n_s]$ is defined as

$$E_{v,B}[n,n_s] = F[n,n_s] + \int n(\bar{r})v(\bar{r})\,d\bar{r} - \int n_s(\bar{r})\beta_e B(\bar{r})\,d\bar{r} \quad , \qquad (1)$$

with

$$F[n,n_s] = \min_{\psi \to n,n_s}\left\{\left\langle \psi \middle| \hat{T} + \hat{V}_{ee} \middle| \psi \right\rangle\right\} . \qquad (2)$$

A ground state corresponding to a given $\left(v(\bar{r}), B(\bar{r})\right)$ can be obtained by the minimization of $E_{v,B}[n,n_s]$ under the conservation constraint of the total electron number

$$N = \int n(\bar{r})\,d\bar{r} \quad . \qquad (3)$$

This leads to the Euler-Lagrange equations

$$\frac{\delta E_{v,B}[n,n_s]}{\delta n(\bar{r})} = \mu \qquad (4)$$

and

$$\frac{\delta E_{v,B}[n,n_s]}{\delta n_s(\bar{r})} = 0 \qquad (5)$$

for the ground-state $\left(n(\bar{r}), n_s(\bar{r})\right)$. Carrying out the differentiations for the known, external part of $E_{v,B}[n,n_s]$ in the above equations,

$$\frac{\delta F[n,n_s]}{\delta n(\bar{r})} + v(\bar{r}) = \mu \qquad (6)$$

and

$$\frac{\delta F[n,n_s]}{\delta n_s(\bar{r})} - \beta_e B(\bar{r}) = 0 \quad . \qquad (7)$$

Because the same ground state $(n(\vec{r}), n_s(\vec{r}))$ is yielded by magnetic fields differing by a constant (i.e., Eq.(7) should hold also with $B(\vec{r}) + \Delta B$), a contradiction arises. This indicates that $F[n, n_s]$ is not differentiable with respect to $n_s(\vec{r})$.

Now, the question arises whether the nondifferentiability of $E_{v,B}[n, n_s]$ with respect to $n_s(\vec{r})$ is a derivative discontinuity, similar to the case of $E_v[n]$ in spin-free DFT [40,41], or $E_{v,B}[n, n_s]$'s derivative with respect to $n_s(\vec{r})$ exists only over domains of constant spin number $N_s = \int n_s(\vec{r})d\vec{r}$. In the case of a derivative discontinuity, one has two derivatives, one for variations of $n_s(\vec{r})$ with $dN_s = \int \delta n_s(\vec{r})d\vec{r} \ge 0$, and one for variations with $dN_s \le 0$, with the two derivatives differing by an $\vec{r}$-independent constant. If the derivative is only defined for the subspace of variations with $\int \delta n_s(\vec{r})d\vec{r} = 0$, it is then a restricted derivative, determined only up to an additive constant [42].

The fact that the ambiguity of $B(\vec{r})$ by an arbitrary constant (which is the only possible ambiguity of $B(\vec{r})$ [30]; see also [34,36,38]) does not conflict with the existence of the restricted derivative $\left. \dfrac{\delta E_{v,B}[n, n_s]}{\delta n_s(\vec{r})} \right|_{N_s}$ (restricted to the domain of $\int n_s(\vec{r})d\vec{r} = N_s$) has been pointed out in [34,43]. The problem due to the ambiguity of $B(\vec{r})$ is avoided if the minimization of $E_{v,B}[n, n_s]$ under the conservation of Eq.(3) is split into two steps. First minimize $E_{v,B}[n, n_s]$ under the additional constraint of fixation of the spin number

$$N_s = \int n_s(\vec{r})d\vec{r} \ , \tag{8}$$

beside Eq.(3); then continue by minimizing with respect to $N_s$. In the first step, fixing $N_s$ introduces another Lagrange multiplier, leading to the Euler-Lagrange equation

$$\left. \frac{\delta F[n, n_s]}{\delta n_s(\vec{r})} \right|_{N_s} - \beta_e B(\vec{r}) = \mu_s \ , \tag{9}$$

in the place of Eq.(7). Of course, the full derivative of $E_{v,B}[n, n_s]$ with respect to $n_s(\vec{r})$ remains nonexisting, but the derivative can now be restricted to the domain of $n_s(\vec{r})$'s with a given $N_s$; for this derivative, the ambiguity of $B(\vec{r})$ then does not lead to a contradiction.

In Eq.(9), $\mu_s$ is determined only after the constant ambiguity in the derivative is fixed by some choice. One may fix the asymptotic value of the derivative explicitly, like the fixation $v_\sigma^{KS}(\infty) = 0$ of the Kohn-Sham spin potentials to obtain a SDFT analog of Koopmans' theorem for the Kohn-Sham energies [44,45]. Alternatively, one can choose $\left.\dfrac{\delta F[n,n_s]}{\delta n_s(\vec{r})}\right|_{N_s}$ to be the number-conserving derivative $\dfrac{\delta F[n,n_s]}{\delta_{N_s} n_s(\vec{r})}$ [46,47] (see [48-50] for applications of

number-conserving functional differentiation). The necessary fixation of $\left.\dfrac{\delta F[n,n_s]}{\delta n_s(\vec{r})}\right|_{N_s}$ to have a unique $\mu_s$ is similar to how, in the Schrödinger equation, the ambiguity of the scalar external potential $v(\vec{r})$ needs to be fixed to have a unique energy.

The appearance of a second Lagrange constant is not strange in SDFT. Looking for the

lowest-energy state with spin number $N_s$ for a given $(v(\vec{r}), B(\vec{r}))$, the energy functional

$E_{v,B}[n,n_s]$ has to be minimized subject to conservation constraint both on $N$ (Eq.(3)) and $N_s$

(Eq.(8)), yielding the Euler-Lagrange equation

$$\frac{\delta E_{v,B}[n,n_s]}{\delta n_s(\vec{r})} = \mu_s \qquad (10)$$

beside Eq.(4). (By construction, the density functional in Eq.(2) gives the correct value for the sum of the kinetic and the interelectron repulsion energy for *every* pair of $n(\vec{r})$ and $n_s(\vec{r})$ that belongs to the lowest-energy state of some spin multiplicity. This means that SDFT is automatically formulated for the lowest-energy state(s) of any $N_s$, not just for ground states.

That is, SDFT is not a "ground-state theory".) The natural emergence of a $\mu_s$, however, does

not eliminate the nondifferentiability of $F[n,n_s]$ arising from the nonuniqueness of $B(\vec{r})$:

because a ground state *can always* be obtained by minimizing $E_{v,B}[n,n_s]$ under only the constraint Eq.(3), $\mu_s$ in Eq.(10) would be uniquely zero for ground states if $F[n,n_s]$ were fully differentiable.

### III. Existence of one-sided derivatives of $F[n,n_s]$ with respect to $n_s(\vec{r})$

We now turn to the question whether $F[n,n_s]$ can have one-sided derivatives. At first sight, one might expect that the one-sided derivatives do not exist either, expecting that if,

say, the right-side derivative of $F[n,n_s]$ existed, then Eq.(5) (or Eq.(7)) could be written as

$$\left. \frac{\delta E_{v,B}[n,n_s]}{\delta n_s(\vec{r})} \right|_{N_s+} = 0 \ , \tag{11}$$

or

$$\left. \frac{\delta F[n,n_s]}{\delta n_s(\vec{r})} \right|_{N_s+} - \beta_e B(\vec{r}) = 0 \ , \tag{12}$$

one-sided derivatives being uniquely determined by their definition. Consequently, the contradiction with the known nonuniqueness of $B(\vec{r})$ would remain.

However, although the right-side derivative is indeed uniquely determined by

$$E_{v,B}[n,n_s + \delta_{N_s+} n_s(\vec{r})] - E_{v,B}[n,n_s] = \int \left. \frac{\delta E_{v,B}[n,n_s]}{\delta n_s(\vec{r})} \right|_{N_s+} \delta_{N_s+} n_s(\vec{r}) d\vec{r} \ , \tag{13}$$

where $\delta_{N_s+} n_s(\vec{r})$ is an *arbitrary* first-order variation with $\int \delta_{N_s+} n_s(\vec{r}) d\vec{r} \geq 0$, it does not follow from the energy variational principle for the variations of $n_s(\vec{r})$ that

$$\left( E_{v,B}[n,n_s + \delta_{N_s+} n_s(\vec{r})] - E_{v,B}[n,n_s] = \right) \int \left. \frac{\delta E_{v,B}[n,n_s]}{\delta n_s(\vec{r})} \right|_{N_s+} \delta_{N_s+} n_s(\vec{r}) d\vec{r} = 0 \ . \tag{14}$$

This is because the full derivative $\dfrac{\delta E_{v,B}[n,n_s]}{\delta n_s(\vec{r})}$ does not exist; therefore the minimum of $E_{v,B}[n,n_s]$ is not a stationary point. Instead, a constant appears on the right of Eq.(11) and Eq.(12),

$$\left. \frac{\delta F[n,n_s]}{\delta n_s(\vec{r})} \right|_{N_s+} - \beta_e B(\vec{r}) = \mu_s^+ \ , \tag{15}$$

just as in the case of the ambiguous $\left. \dfrac{\delta E_{v,B}[n,n_s]}{\delta n_s(\vec{r})} \right|_{N_s}$, since the (first-order) energy variations still vanish over the domain of $\int n_s(\vec{r}) d\vec{r} = N_s$. It is interesting to recognize that the uniquely determined one-sided derivatives behave in a minimization just like the $N$-restricted derivatives. The difference will be that the Lagrange constant $\mu_s$ is now fixed by the extension from the fixed-$N_s$ domain: as $\mu_s^+$, or $\mu_s^-$. (Note that a difference between the two one-sided derivatives would not be allowed by Eq.(12), since Eq.(12) could be written with the left-side derivative too. That is, Eq.(12) would *itself* say, irrespective of $B(\vec{r})$'s nonuniqueness, that the one-sided derivatives exist only if the full derivative of $F[n,n_s]$ exists.)

Before ending this section, an important exception to the nonexistence of the full derivative $\dfrac{\delta E_{v,B}[n, n_s]}{\delta n_s(\vec{r})}$ for ground states should be emphasized, namely, the case of degenerate ground states with different spin numbers. Consider for example the ground-state Li atom with $B(\vec{r}) = 0$. Between $N_s = -1$ and $N_s = 1$, the energy is constant with respect to $N_s$, meaning there should be no nondifferentiability problem for $n_s(\vec{r})$'s with $N_s$ in that interval. This may seem to contradict the nonuniqueness of $B(\vec{r})$; however, notice that one can shift $B(\vec{r})$ by a constant without a change of the ground state *only as long as* a crossing of energy levels does not occur [30,31], while the considered case is *just at* a level crossing. Switching on a (constant) magnetic field in the up/down direction will cause the energy of the state with $N_s = 1 / N_s = -1$ go below the energy of any of the states with $-1 < N_s < 1$. Consequently, apart from the level crossing at $B(\vec{r}) = 0$, Li does not have another ground state with $-1 < N_s < 1$, meaning that there is no nonuniqueness of $B(\vec{r})$ in this case. As another example, with $B(\vec{r}) \neq 0$, the Be atom can be mentioned (see Fig.1(a) of [30] for an illustration of its level crossings). Be has a degenerate ground state at $B(\vec{r}) = \pm(E_1 - E_0)/(2\beta_e)$, e.g., with $E_1$ and $E_0$ being the energy of the first excited state ($1s^2 2s 2p$, with $N_s = \pm 2$) and the energy of the ground state ($1s^2 2s^2$, with $N_s = 0$), respectively, without a magnetic field. This level crossing is a mixture of an $N_s = 0$ and an $N_s = \pm 2$ state. (For illustration, see Figure 1, with a mixed spin-number state explicitly displayed.) In summary, $B(\vec{r})$'s nonuniqueness does not cause a derivative discontinuity for degenerate ground states that are mixtures of two different spin number states ($N_s'$ and $N_s''$), for $N_s' < N_s < N_s''$, since there is actually no nonuniqueness in that case.

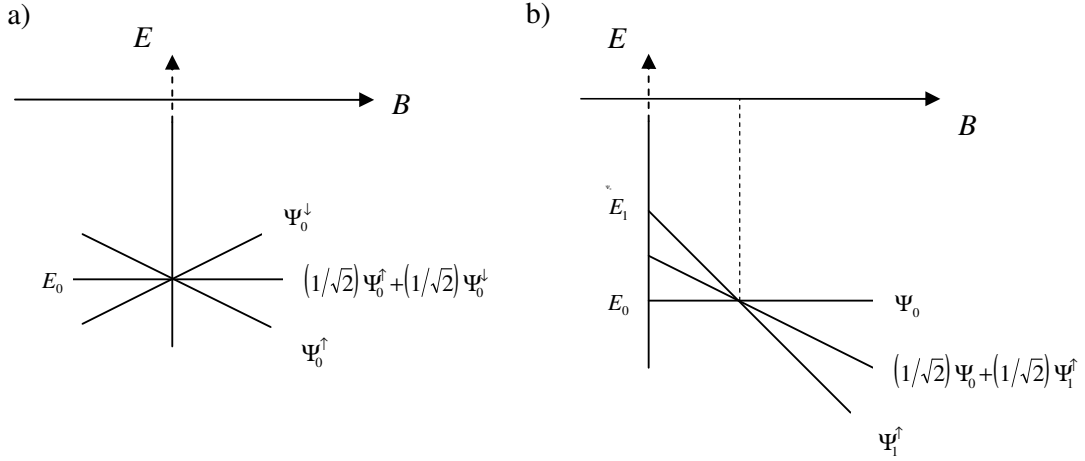

FIG. 1. Level crossing of (a) the Li atom at $B = 0$ and (b) the Be atom at $B = (E_1 - E_0)/(2\beta_e)$, explicitly displaying the state that is the 50%-50% mixture of (a) the ground states $(\uparrow\downarrow,\uparrow)$ and $(\uparrow\downarrow,\downarrow)$ (at $B = 0$) and (b) the ground state $(\uparrow\downarrow,\uparrow\downarrow)$ and the first excited state $(\uparrow\downarrow,\uparrow,\uparrow)$ (at $B = 0$).

## IV. Identification of $\mu_s^+$ and $\mu_s^-$ as one-sided energy derivatives with respect to the spin number

We thus showed that the nonuniqueness of the external magnetic field $B(\bar{r})$ does not exclude the existence of one-sided derivatives of $F[n, n_s]$ with respect to $n_s(\bar{r})$. It will now be demonstrated that the Lagrange multipliers $\mu_s^+$ and $\mu_s^-$ can be identified with the right-side and the left-side derivative with respect to $N_s$, respectively, of the energy $E[N, N_s, v, B]$ of the given $N$-electron system, in external fields $v(\bar{r})$ and $B(\bar{r})$. Note that without the existence of one-sided derivatives, the Lagrange multiplier $\mu_s$ of Eq.(9) could not be interpreted as a derivative of $E[N, N_s, v, B]$ with respect to $N_s$, i.e., as a (spin) chemical potential, leading to a breakdown of the thermodynamical analogy of density functional theory [51-53] in its spin-polarized extension. Of course, due to the ambiguity of $\left.\dfrac{\delta F[n, n_s]}{\delta n_s(\bar{r})}\right|_{N_s}$ in Eq.(9), $\mu_s$ could be fixed as $\left.\dfrac{\partial E[N, N_s, v, B]}{\partial N_s}\right|_+$, e.g., but this would only be a formal

enforcement of the thermodynamical analogy, and $\left.\dfrac{\partial E[N, N_s, v, B]}{\partial N_s}\right|_+$ still could not be obtained as the (right-side) derivative of $E_{v,B}[n, n_s]$ with respect to the spin density, consequently would not be defined as a density functional. It is interesting to recognize though that even if it were impossible to identify $\mu_s^{+(-)}$ as $\left.\dfrac{\partial E[N, N_s, v, B]}{\partial N_s}\right|_{+(-)}$, it would not be as tragic for conceptual DFT as might be expected. The definitions of most of the reactivity descriptors do not require differentiation with respect to the density, or spin density, and even do not involve density functional theory. Most of the reactivity indicators are defined as the (first- or higher-order) derivatives of $E[N, N_s, v, B]$ with respect to its arguments. A quantity whose SDFT generalization would be affected by the breakdown of the thermodynamical analogy is the local hardness [54,55], being defined as $\eta(\bar{r}) = \left(\dfrac{\delta \mu}{\delta n(\bar{r})}\right)_{v(\bar{r})}$ ($\mu$ denoting here the derivative of the energy $E[N, v]$ with respect to $N$). Note, however, that defining the local hardness is problematic even in spin-free DFT [56-62].

To show that $\mu_s^+$ in Eq.(15) equals $\left.\dfrac{\partial E[N, N_s, v, B]}{\partial N_s}\right|_+$, we will generalize the procedure Parr and Bartolotti [53] used, in the spin-independent case, to derive $\mu = \dfrac{\partial E[N, v]}{\partial N}$, assuming the existence of the full derivatives $\dfrac{\delta E_v[n]}{\delta n(\bar{r})}$ and $\dfrac{\partial E[N, v]}{\partial N}$. Their argument was extended by Galván et al. [8] for spin-polarized DFT, but without addressing the effects of the nonuniqueness of $B(\bar{r})$.

The first-order change in the energy $E[N, N_s, v, B]$ induced by a transition from a minimum-energy state of some spin number to another one with the same number of electrons can be written as

$$\delta_N E[N, N_s, v, B] = \left.\frac{\partial E[N, N_s, v, B]}{\partial N_s}\right|_+ \partial_+ N_s$$

$$+ \int \frac{\delta E[N, N_s, v, B]}{\delta v(\bar{r})} \delta v(\bar{r}) d\bar{r} + \int \frac{\delta E[N, N_s, v, B]}{\delta B(\bar{r})} \delta B(\bar{r}) d\bar{r} \quad . \quad (16)$$

On the other hand, the same energy variation is given by

$$\delta_N E_{v,B}[n,n_s] = \int \frac{\delta E_{v,B}[n,n_s]}{\delta n(\bar{r})}\bigg|_N \delta_N n(\bar{r}) d\bar{r} + \int \frac{\delta E_{v,B}[n,n_s]}{\delta n_s(\bar{r})}\bigg|_{N_s+} \delta_{N_s+} n_s(\bar{r}) d\bar{r}$$

$$+ \int \frac{\delta E_{v,B}[n,n_s]}{\delta v(\bar{r})} \delta v(\bar{r}) d\bar{r} + \int \frac{\delta E_{v,B}[n,n_s]}{\delta B(\bar{r})} \delta B(\bar{r}) d\bar{r} \quad . \tag{17}$$

The existence of the derivatives in both of the above equations is assumed. In Eq.(17), the first term on the right cancels out, due to (i) the constancy of the derivative inside the integrand (following from the corresponding Euler-Lagrange equation), and (ii) $\int \delta_N n(\bar{r}) d\bar{r} = 0$. Further, from Eq.(1),

$$\frac{\delta E_{v,B}[n,n_s]}{\delta v(\bar{r})} = n(\bar{r}) \tag{18}$$

and

$$\frac{\delta E_{v,B}[n,n_s]}{\delta B(\bar{r})} = -\beta_e\, n_s(\bar{r}) \quad . \tag{19}$$

Comparing Eqs.(16) and (17), and utilizing the arbitrariness of $\partial_+ N_s$, $\delta v(\bar{r})$, and $\delta B(\bar{r})$,

$$\mu_s^+ = \frac{\partial E[N,N_s,v,B]}{\partial N_s}\bigg|_+ \quad , \tag{20}$$

$$\frac{\delta E[N,N_s,v,B]}{\delta v(\bar{r})} = n(\bar{r}) \quad , \tag{21}$$

and

$$\frac{\delta E[N,N_s,v,B]}{\delta B(\bar{r})} = -\beta_e\, n_s(\bar{r}) \tag{22}$$

are obtained, where use of Eq.(15), and $\int \delta_{N_s+} n_s(\bar{r}) d\bar{r} = \partial_+ N_s$ is made. (In addition to Eq.(20), it also follows thus that, if the derivatives in Eqs.(16) and (17) exist, the energy derivatives with respect to $v(\bar{r})$ and $B(\bar{r})$ are $n(\bar{r})$ and $-\beta_e\, n_s(\bar{r})$, respectively, in accordance with the functional generalization of the Hellmann-Feynman theorem.) $\mu_s^-$ can of course be similarly identified as

$$\mu_s^- = \frac{\partial E[N,N_s,v,B]}{\partial N_s}\bigg|_- \quad . \tag{23}$$

In the above, the nondegeneracy of the considered state is assumed, because for degenerate states the energy derivatives with respect to the external fields do not exist. This is not a problem, however, because to obtain Eqs.(20) and (23) it is sufficient to consider energy variations in fixed external fields. Therefore the last two terms in both of Eqs.(16) and (17) can be omitted, avoiding the need to require nondegeneracy. (Another possibility is to apply the generalization [63] of the Hellmann-Feynman theorem to degenerate states.)

We wish to emphasize that Eqs.(20) and (23) are valid not only for ground states, but for the lowest-energy state of any spin multiplicity. Giving a fractional electron number generalization of $E(N)$ [40,41,64], the energy variations of Eqs.(16) and (17) can also include $N$-changing variations, so that $\mu^{(+/-)}$ can be identified with the energy derivative with respect to $N$. With Eq.(20) (or Eq.(23)), it is easily seen that the Lagrange multiplier $\mu_s^+$ (or $\mu_s^-$) indeed shifts in accordance with $B(\bar{r})$ in Eq.(15), since from the Schrödinger equation,

$$E[N, N_s, v, B + \Delta B] = E[N, N_s, v, B] - N_s \beta_e \Delta B \quad , \tag{24}$$

which then gives

$$\mu_s^+[N, N_s, v, B + \Delta B] = \mu_s^+[N, N_s, v, B] - \beta_e \Delta B \quad . \tag{25}$$

This is analogous to the shift of $\mu^{(+/-)}$ with $v(\bar{r})$.

Note that the nonexistence of the full derivative $\dfrac{\delta F[n, n_s]}{\delta n_s(\bar{r})}$ implies the nonexistence of $\dfrac{\partial E[N, N_s, v, B]}{\partial N_s}$; that is, $\dfrac{\partial E[N, N_s, v, B]}{\partial N_s}\bigg|_+ \neq \dfrac{\partial E[N, N_s, v, B]}{\partial N_s}\bigg|_-$. Otherwise $\mu_s^+$ would be equal to $\mu_s^-$, yielding $\dfrac{\delta F[n, n_s]}{\delta n_s(\bar{r})}\bigg|_{N_s+} = \dfrac{\delta F[n, n_s]}{\delta n_s(\bar{r})}\bigg|_{N_s-}$, which would just mean the existence of a full derivative. It is worth pointing out that this argument is the opposite of what is done in the case of the fractional particle number generalization of $E_v[n]$ given in [40], where the discontinuity of $E_v[n]$'s derivative (at integer $N$) is inferred from the discontinuity of $E[N, v]$'s derivative with respect to $N$. Of course, in the case of $E(N_s)$ too, where a (minimum) energy value is naturally determined by the Schrödinger equation with the given $v(\bar{r})$ and $B(\bar{r})$ for fractional spin numbers $N_s$ as well, the discontinuity with respect to $N_s$ of the energy derivative [64,65] can be established without recourse to density functional theory. However, in the case of $E(N_s)$, the nonuniqueness of $B(\bar{r})$ itself leads to a discontinuity in $\dfrac{\partial E[N, N_s, v, B]}{\partial N_s}$, via density functional theory, while the ambiguity of $v(\bar{r})$ is neutralized by the need to specify the electron number in the minimization of the energy functional $E_v[n]$.

The conclusions of this study also apply to the $(N_\uparrow, N_\downarrow)$ representation of SDFT. In that case,

$$\left.\frac{\delta F[n_\uparrow, n_\downarrow]}{\delta n_\uparrow(\vec{r})}\right|_{N_\uparrow+} + v(\vec{r}) - \beta_e B(\vec{r}) = \mu_\uparrow^+ \ , \qquad (26)$$

$$\left.\frac{\delta F[n_\uparrow, n_\downarrow]}{\delta n_\downarrow(\vec{r})}\right|_{N_\downarrow+} + v(\vec{r}) + \beta_e B(\vec{r}) = \mu_\downarrow^+ \ , \qquad (27)$$

with

$$\mu_\uparrow^+ = \left.\frac{\partial E[N_\uparrow, N_\downarrow, v, B]}{\partial N_\uparrow}\right|_+ \ , \qquad (28)$$

$$\mu_\downarrow^+ = \left.\frac{\partial E[N_\uparrow, N_\downarrow, v, B]}{\partial N_\downarrow}\right|_+ \ . \qquad (29)$$

(Both Eqs.(26) and (27) can be written with a – instead of the + in the indices as well.) Note that to have the one-sided derivatives in the above equations, a proper fractional particle number generalization for the energy is necessary. By contrast, in the $(N, N_s)$ representation, the $N$ dependence is separated from the spin dependence. This latter fact was critical for showing that the nonuniqueness of $B(\vec{r})$ induces a derivative discontinuity in the energy density functional, since the zero-temperature grand canonical ensemble extension [40] of the energy for fractional particle number (which is the only physical extension [40,41]) already yields a derivative discontinuity of $F[n_\uparrow, n_\downarrow]$ at integer spin numbers $N_\uparrow$ and $N_\downarrow$ (even for the $N_s = 0$ ground state of the Li atom!). This shows that although the $(N_\uparrow, N_\downarrow)$ representation may be more attractive intuitively, it mixes the two kinds of derivative discontinuities. It seems useful to separate these discontinuities when developing accurate approximate density functionals, where accounting for the discontinuities is crucial [66].

## V. Conclusions

In conclusion, the thermodynamical analogy [51,8,11] of density functional theory is not compromised by the nonuniqueness [30,31] of the external magnetic field $B(\vec{r})$ in spin-polarized DFT. It is shown that the ambiguity of $B(\vec{r})$, though excludes the existence of a full derivative of the energy density functional $E_{v,B}[n, n_s]$ with respect to $n_s(\vec{r})$, does not exclude the existence of the one-sided derivatives. This ensures the mathematical rigor of the spin-resolved reactivity descriptors, and their connections to density functional derivatives, which is an essential element in the thermodynamical interpretation of DFT [53]. In particular, the one-sided derivatives of $E_{v,B}[n, n_s]$ with respect to the spin density are equal to the one-sided energy derivatives with respect to the spin number. This makes the concept of the spin chemical potential well-established. Because of the derivative discontinuity in $N_s$, separate reactivity indices have to be used for processes with an increase in the spin number, and for processes with a decrease in it. Together with the well-known discontinuity with respect to the electron number, this induces a multiplication of the chemical reactivity descriptors. It is important to emphasize, however, that conceptually no new indices emerge in this way.

**Acknowledgments:** T.G. acknowledges a Visiting Postdoctoral Fellowship from the Fund for Scientific Research – Flanders (FWO). P.W.A. acknowledges support from NSERC and Sharcnet.